\documentstyle[psfig]{mn} 
\title{The old open clusters Saurer A, B and C revisited\thanks{Based 
on observations  
carried out at ESO La Silla. All the photometry is available 
at WEBDA database: http://obswww.unige.ch/webda/navigation.html}} 
\author[Carraro \& Baume]        
{Giovanni Carraro$^1$ and  Gustavo Baume$^{1,2}$ 
\thanks{email: 
giovanni.carraro@unipd.it (GC); baume@pd.astro.it (GB)}\\ 
$^1$Dipartimento di Astronomia, Universit\`a di Padova, Vicolo 
dell'Osservatorio 2, I-35122 Padova, Italy \\
$^2$Facultad de Ciencias Astron\'omicas y Geof\'{\i}sicas de la UNLP,
IALP-CONICET, Paseo del Bosque s/n, La Plata, Argentina \\
 } 
 
\date{\it Submitted: June 2003} 
\pubyear{2003} 
\begin{document} 
\maketitle 
\title{The open clusters Saurer A, B and C} 
 
\begin{abstract} 
We report on deep (V $\approx$ 24.0) $VI$ CCD photometry of 
3 fields centered in the regions of the old
open clusters Saurer A, B and C. In the case of Saurer A,
which is considered one of the oldest known open cluster,
we also provide a comparison field. From the analysis
of the photometry we claim that Saurer A is as old as M 67
($\approx$ 5 Gyrs),
but more metal poor (Z=0.008). Moreover it turns out to be
the open cluster with the largest galactocentric distance
so far detected.\\ 
As for Saurer B, it closely resembles NGC~2158,
and indeed is  of intermediate-age (1.8-2.2 Gyrs)
and significantly reddened. In this case we revise both the age and 
the distance with respect to previous studies, but we are not able to
clearly establish the cluster metal abundance.\\
Finally, Saurer C has an age of about 2 Gyrs, but we emphasize
that  the precise determination of its properties is hampered by the heavy field
stars contamination.
\end{abstract} 
 

\section{Introduction} 
Saurer et al. (1994) identified 6 star concentrations by 
inspecting POSS or ESO/SERC atlas, which they suggest might
represent hitherto uncatalogued star clusters, which
to date have not been included in any open clusters
catalog (Dias et al. 2002).\\
Preliminary photometry of all these star concentrations
have been recently published by Frinchaboy \& Phelps (2002, FP02 hereinafter).
Their results can be summarized as follows:

\begin{description} 
\item $\bullet$ Saurer A, B and C are old open clusters
with ages greater than 2.5 Gyr. In particular Saurer~A
is marked as a very promising target for further studies
due to the combination of his very large distance and age; 
\item $\bullet$ Saurer E is probably not a physical cluster;
\item $\bullet$  Saurer D and F finally are intermediate age
open cluster, with ages between 1 and 2 Gyr.
\end{description} 
  
Due to the relevant importance of the oldest open clusters
for our understanding of the formation and early evolution
of the Galactic disk (Janes \& Phelps 1994, 
Carraro \& Chiosi 1994, Friel 1995, Carraro et al. 1998, 
Bragaglia et al. 2000), 
we decided to obtain new, better quality
and deeper photometry of the oldest clusters in this sample,
namely Saurer A, B and C (see Table~1), to better constrain their
fundamental parameters.\\ 
Moreover, instead of deriving ages from a so-called Morphological
Age Indicator (MAI, Janes \& Phelps 1994), 
we are presenting a different approach to the determination
of all the clusters basic parameters, which is based on the
comparison of the photometry with theoretical models.\\

\noindent
The plan of the paper is as follows. Sect.~2 illustrates  
the observation and reduction strategies. 
Sections 3 to 5 are dedicated to the discussion 
of our data for Saurer A, B and C, respectively.
In these Sections we derive estimates of radii, distances,
ages and reddenings.
Finally, Sect.~6 summarizes our findings.

\begin{table}
\caption{{}Basic data of the observed objects.}
\fontsize{8} {10pt}\selectfont
\begin{tabular}{ccccc}
\hline
\multicolumn{1}{c} {$Name$} &
\multicolumn{1}{c} {$\alpha_{2000}$}  &
\multicolumn{1}{c} {$\delta_{2000}$}  &
\multicolumn{1}{c} {$l$} &
\multicolumn{1}{c} {$b$} \\
\hline
Saurer A & 07:20:56 & +01:48:29 & $214.61^{\circ}$ & $+7.21^{\circ}$ \\
Field    & 07:18:18 & +01:53:43 & $214.31^{\circ}$ & $+6.84^{\circ}$\\
Saurer B & 08:25:28 & -39:38:02 & $257.95^{\circ}$ & $-1.06^{\circ}$ \\     
Saurer C & 10:41:25 & -55:18:20 & $285.05^{\circ}$ & $+2.98^{\circ}$ \\     
\hline
\end{tabular}
\end{table}

\section{Observations and Data Reduction} 
 
$\hspace{0.5cm}$
CCD $VI$ observations were carried out with the new  EMMI read arm
camera on board NTT 
at ESO, La Silla, in the photometric night of 
December 9, 2002 and in sub-arcsec seeing conditions. 
The new camera has a mosaic of two 2048 $\times$ 4096 pixels
CCDs which samples 9.9 $\times$ 9.1 arcmin in the sky having a pixel scale
of $0^{\prime\prime}.332$ ($2 \times 2$ binning). 

\noindent
Details of the observations are listed in Table~2 where the observed 
fields are 
reported together with the exposure times, the typical seeing values and the 
air-masses. Figs.~1 to 4 show the finding charts for Saurer A, B, C 
and the comparison field, respectively,
taken from the DSS-2\footnote{Second generation Digitized Sky Survey,
{\tt http://cadcwww.dao.nrc.ca/cadcbin/getdss}} archive. 
The data has been reduced with the 
IRAF\footnote{IRAF is distributed by NOAO, which are operated by AURA under 
cooperative agreement with the NSF.} 
packages CCDRED, DAOPHOT, and PHOTCAL using the point spread function (PSF)
method (Stetson 1987). The calibration equations obtained by observing Landolt 
(1992) PG 0918+029, SA 098-562, SA 101-424  and PG 0942-029 fields observed
along the night, are:

\begin{figure}  
\centerline{\psfig{file=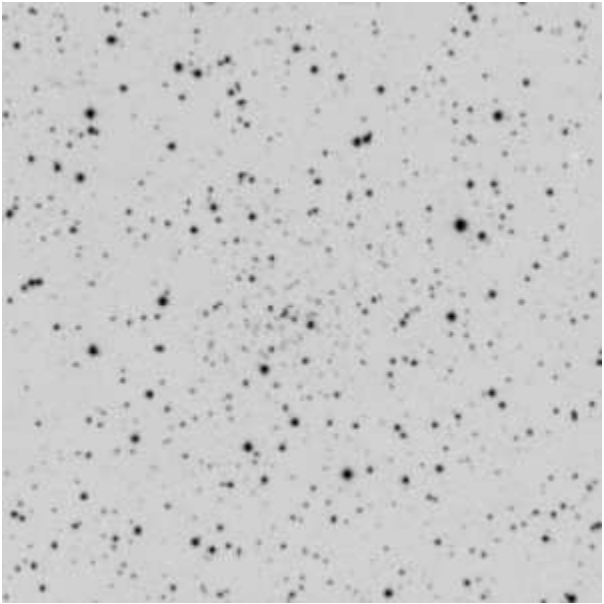,width=8cm,height=8cm}} 
\caption{A DSS image centered on Saurer A. The field of view
is 10 squared arcmin. North is up, East on the left.} 
\label{SaurerA} 
\end{figure}

\begin{figure}  
\centerline{\psfig{file=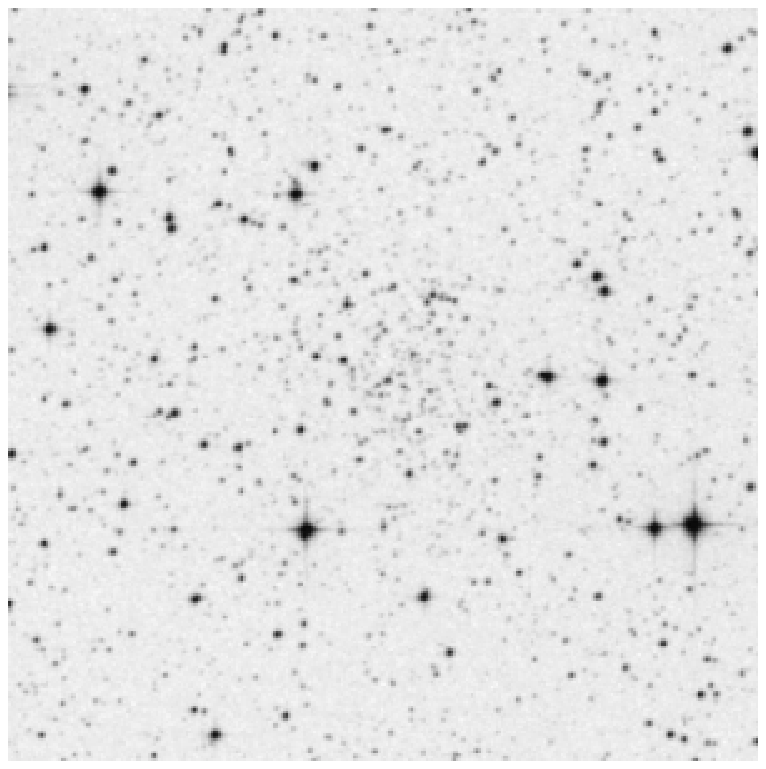,width=8cm,height=8cm}} 
\caption{A DSS image centered on Saurer B. The field of view
is 10 squared arcmin. North is up, East on the left.} 
\label{SaurerB} 
\end{figure} 

\begin{figure}  
\centerline{\psfig{file=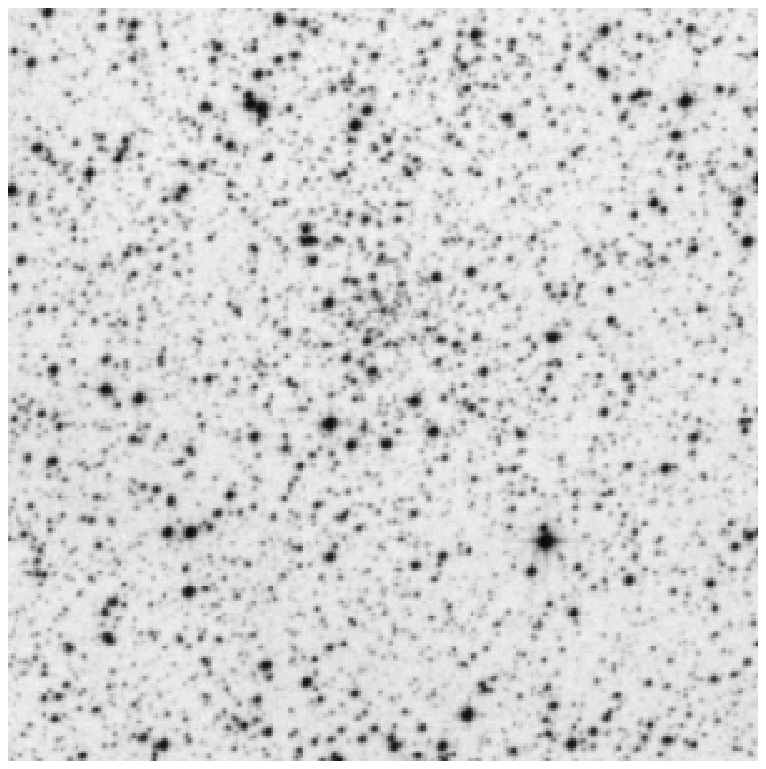,width=8cm,height=8cm}} 
\caption{A DSS image centered on Saurer C. The field of view
is 10 squared arcmin. North is up, East on the left.} 
\label{SaurerC} 
\end{figure} 

\begin{figure}  
\centerline{\psfig{file=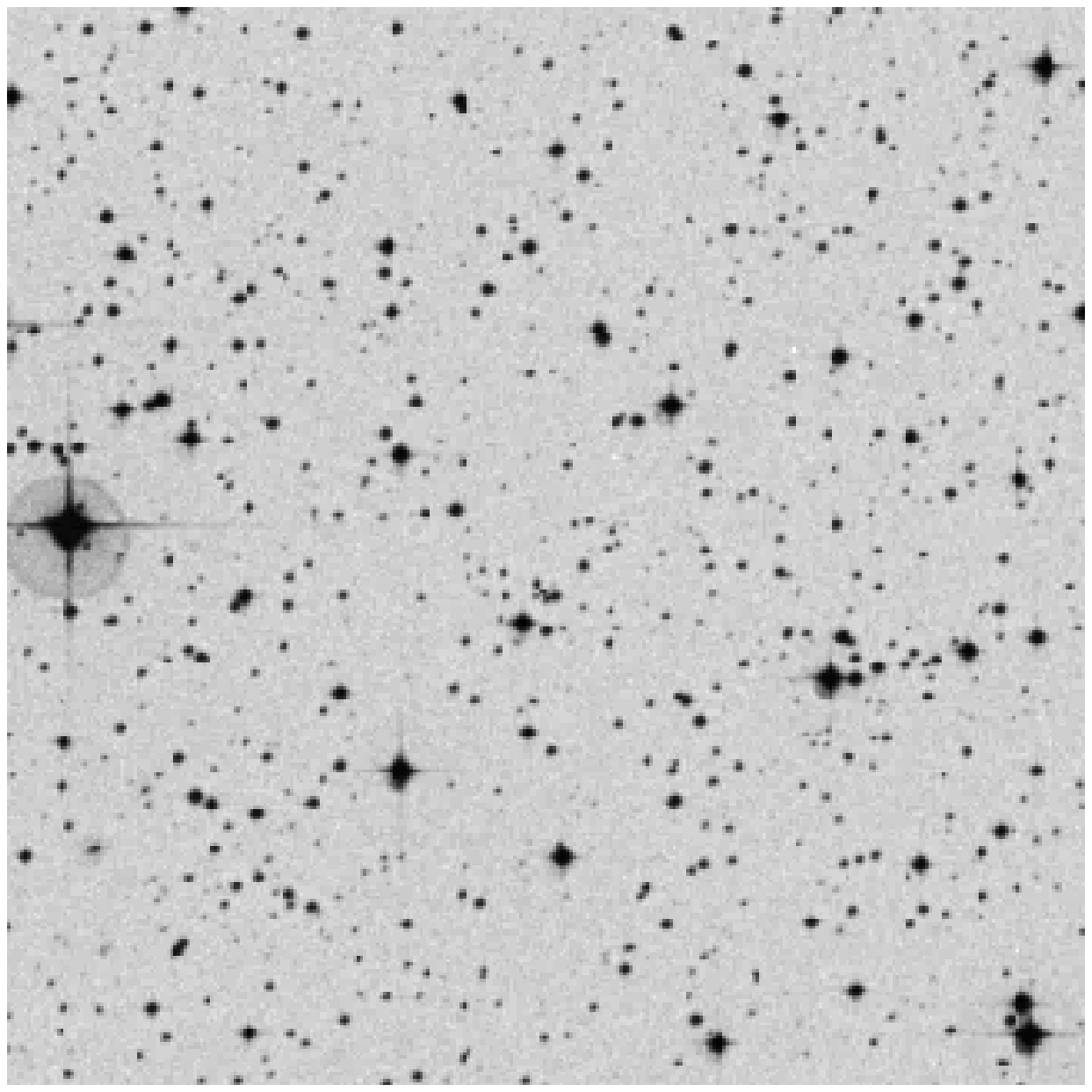,width=8cm,height=8cm}} 
\caption{A DSS image centered the on Saurer A
 comparison field. The field of view
is 10 squared arcmin. North is up, East on the left.} 
\label{field} 
\end{figure}

\begin{table} 
\fontsize{8} {10pt}\selectfont
\caption{Journal of observations of Saurer A, B and C,  and standard star fields 
(December 9, 2002).} 
\begin{tabular}{ccccccc} 
\hline 
\multicolumn{1}{c}{Field}         & 
\multicolumn{1}{c}{Filter}        & 
\multicolumn{3}{c}{Exposure time} & 
\multicolumn{1}{c}{Seeing}        &
\multicolumn{1}{c}{Airmass}       \\
 & & \multicolumn{3}{c}{[sec.]} & [$\prime\prime$] & \\ 
\hline 
 Saurer~A         & V &  1   &  30 &  360 & 0.9 & 1.175 \\ 
                  & I &  1   &  30 &  300 & 0.9 & 1.182 \\

 Saurer~A (Field) & V &  1   &  30 &  360 & 0.9 & 1.180 \\ 
                  & I &  1   &  30 &  300 & 0.9 & 1.188 \\

 Saurer~B         & V &  1   &  30 &  360 & 1.0 & 1.027 \\ 
                  & I &  1   &  30 &  300 & 1.0 & 1.031 \\

 Saurer~C         & V &  1   &  30 &  360 & 0.9 & 1.205 \\ 
                  & I &  1   &  30 &  300 & 0.9 & 1.219 \\
\hline
PG 0918+029       & V &   15   &  &  & 1.0 & 1.278 \\ 
                  & I &   10   &  &  & 1.0 & 1.274 \\ 

PG 0942-029       & V &   15   &  &  & 1.0 & 1.267 \\ 
                  & I &   10   &  &  & 1.0 & 1.263 \\

SA 098-562        & V &   15   &  &  & 0.9 & 1.143 \\ 
                  & I &   10   &  &  & 0.9 & 1.143 \\

SA 101-424        & V &   15   &  &  & 1.0 & 1.148 \\ 
                  & I &   10   &  &  & 1.0 & 1.147 \\ 
\hline
\end{tabular}
\end{table}

\begin{center}
\begin{tabular}{lc}
$v$ = $V$ -0.560$\pm$0.023 -(0.058$\pm$0.023) $\cdot$ $(V-I)$ + 0.135$\cdot$ $X$  \\  
$i$ = $I$ -0.258$\pm$0.066 -(0.063$\pm$0.070) $\cdot$ $(V-I)$ + 0.048$\cdot$ $X$  \\
\end{tabular}
\end{center}

\noindent
where $VI$ are standard magnitudes, $vi$ are the instrumental ones, 
and $X$ is 
the airmass. The standard 
stars in these fields provide a very good color coverage.
For the extinction coefficients, we assumed the 
typical values for La Silla observatory.\\
 
\noindent
The photometry turns out to be quite accurate with global
errors (zero point, PSF fitting and aperture correction errors)
amounting to less than 0.10~mag in magnitude
and 0.15~mag in colour down to V$\approx$ 23.0.\\
The final photometric data are  
available in electronic form at the  
WEBDA\footnote{http://obswww.unige.ch/webda/navigation.html} site. \\

\noindent 
Our photometry extends down to V = 14, and therefore
results to be  about 2 mag deeper than FP02. The photometric accuracy,
excellent seeing conditions and good pixel scale allow
us to properly study very faint and contaminated objects
like those ones in the present study.

\begin{figure}  
\centerline{\psfig{file=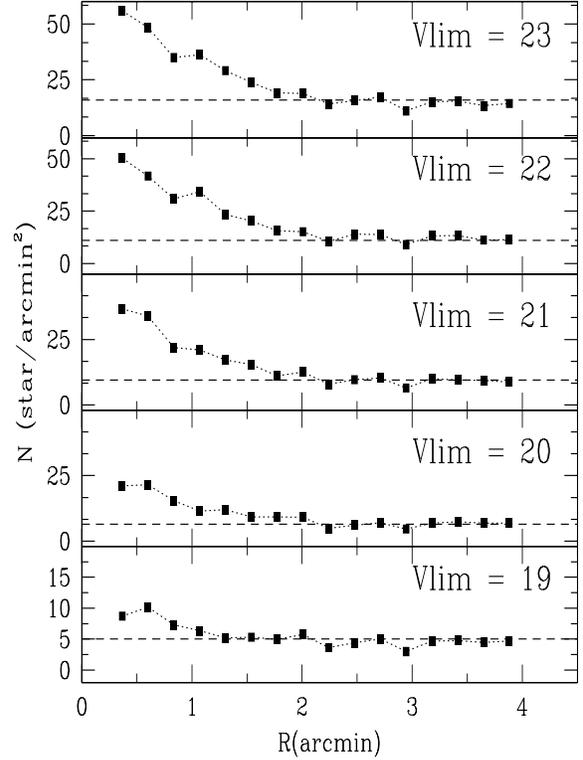,width=8cm,height=11cm}} 
\caption{Star counts in the region of Saurer A as a function
of the V magnitude. The dashed line indicates the level
of the field at the corresponding limiting magnitude.} 
\label{SaurerA} 
\end{figure} 

\section{The open cluster Saurer A}

\subsection{Cluster radius}
Saurer A appears as a very weak concentration of stars (see Fig.~1),
with a size of a few arcmin.
In order to infer a more robust estimate of the cluster radius, we performed
star counts by using our CCD data (1500 stars).
We derived the surface stellar density by performing star counts
in concentric rings  15 arcsecs in size (46 pixels) 
around the visual cluster center,
and then dividing by their respective surfaces.
Our aim is to find the region where the cluster clearly emerges
over the field.
The result is shown in Fig.~5. Here we plot the radial density
profile for the cluster (solid symbols) 
as a function of the limiting magnitude.
We performed star counts 
also in the comparison field, by simply computing the 
density of field stars down to the same limiting magnitude.
The cluster does not emerge much 
from the field when considering
$V_{lim}=19$, which means that the cluster is very poorly populated 
by bright stars. At dimmer magnitudes, the cluster clearly emerges
from the field up to R $\approx$ 1.3 arcmin, that we shall consider
as the cluster radius in the following analysis.

\begin{figure}  
\centerline{\psfig{file=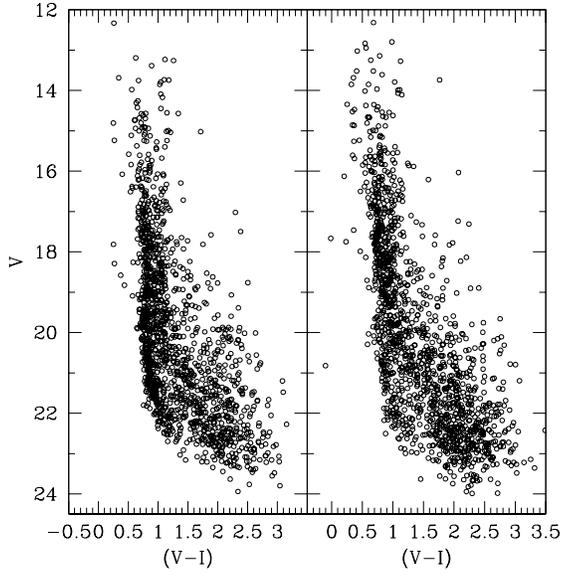,width=8cm,height=8cm}} 
\caption{{\bf Left panel:} CMD of Saurer A.
{\bf Right panel:} CMD of the comparison field.
Here we consider all the measured stars.} 
\label{SaurerA} 
\end{figure}

\begin{figure}  
\centerline{\psfig{file=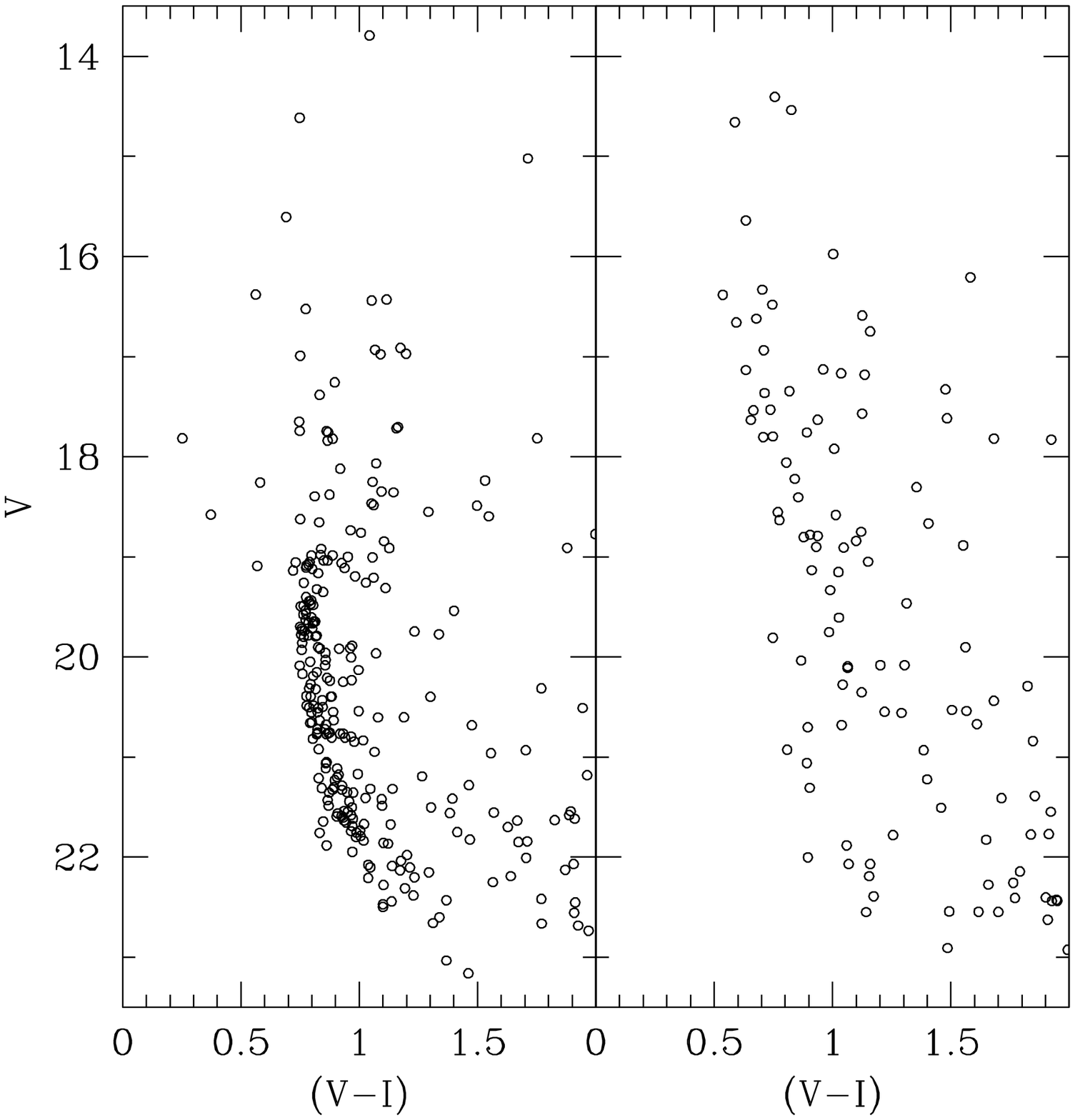,width=8cm,height=8cm}} 
\caption{{\bf Left panel:} CMD of Saurer A.
{\bf Right panel:} CMD of the comparison field.
Here we consider only stars within 1.7 arcmin from the cluster
center.} 
\label{SaurerA} 
\end{figure} 

\begin{figure}  
\centerline{\psfig{file=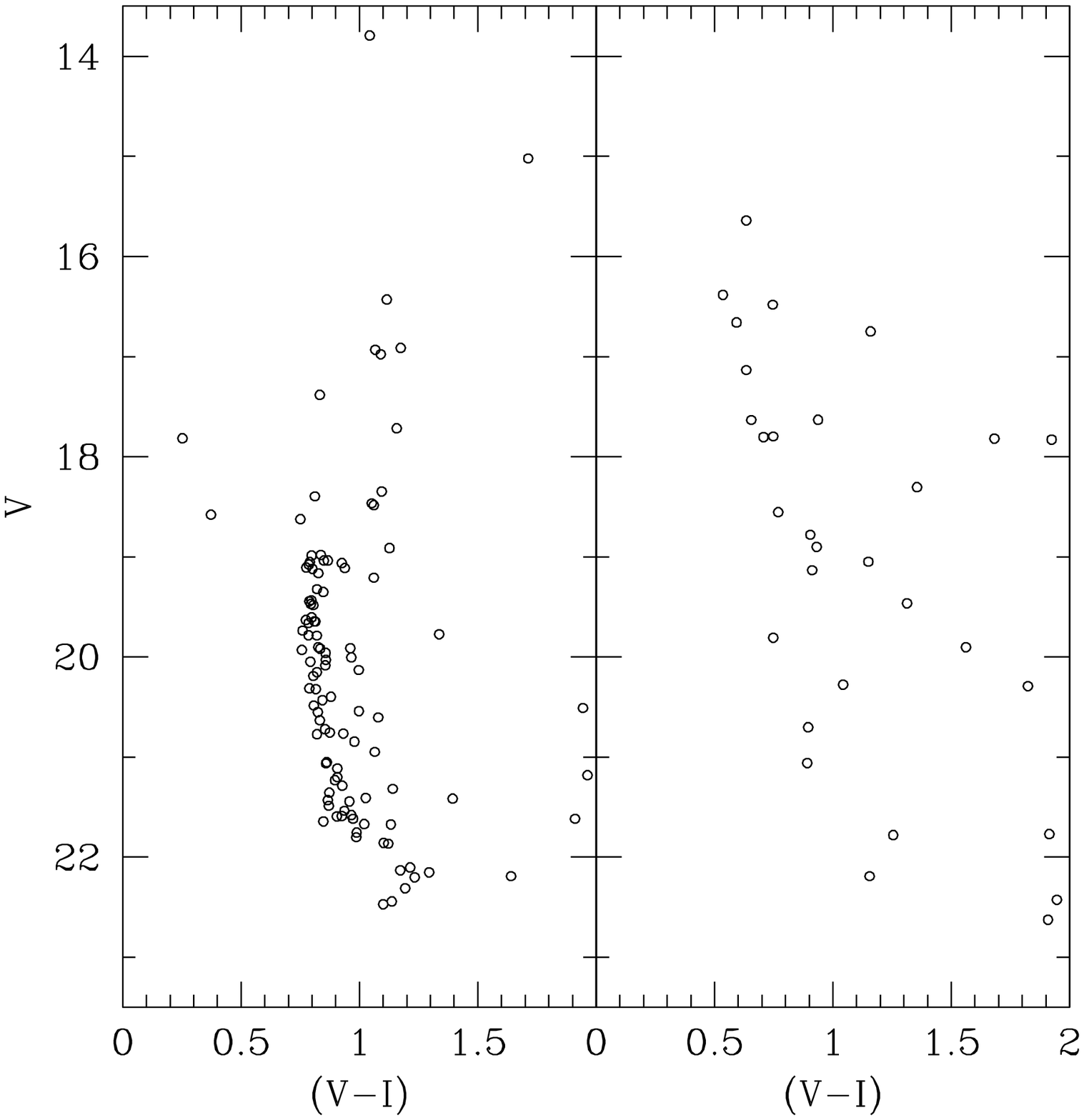,width=8cm,height=8cm}} 
\caption{{\bf Left panel:} CMD of Saurer A.
{\bf Right panel:} CMD of the comparison field.
Here we consider only stars within 0.8 arcmin from the cluster
center.} 
\label{SaurerA} 
\end{figure}

\begin{figure} 
\centerline{\psfig{file=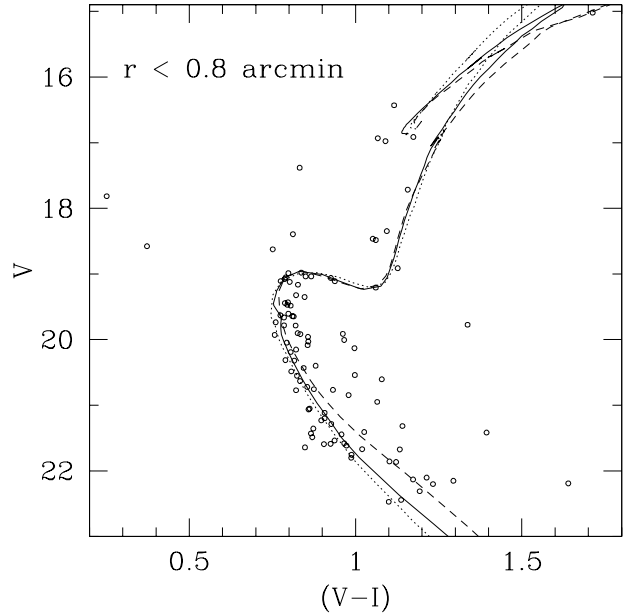,width=\columnwidth}} 
\caption{Age determination for Saurer A
(only stars within 0.8  arcmin from the cluster center
are considered). The {\bf solid line} is a 5 Gyr isochrone
for Z=0.008 metallicity, the {\bf dashed line} is
a 6.3 Gyr isochrone for solar metallicity, and {\bf dotted line}
is a 4.5 Gyr isochrone for Z=0.004.
See text for more details} 
\end{figure} 
 
\begin{figure} 
\centerline{\psfig{file=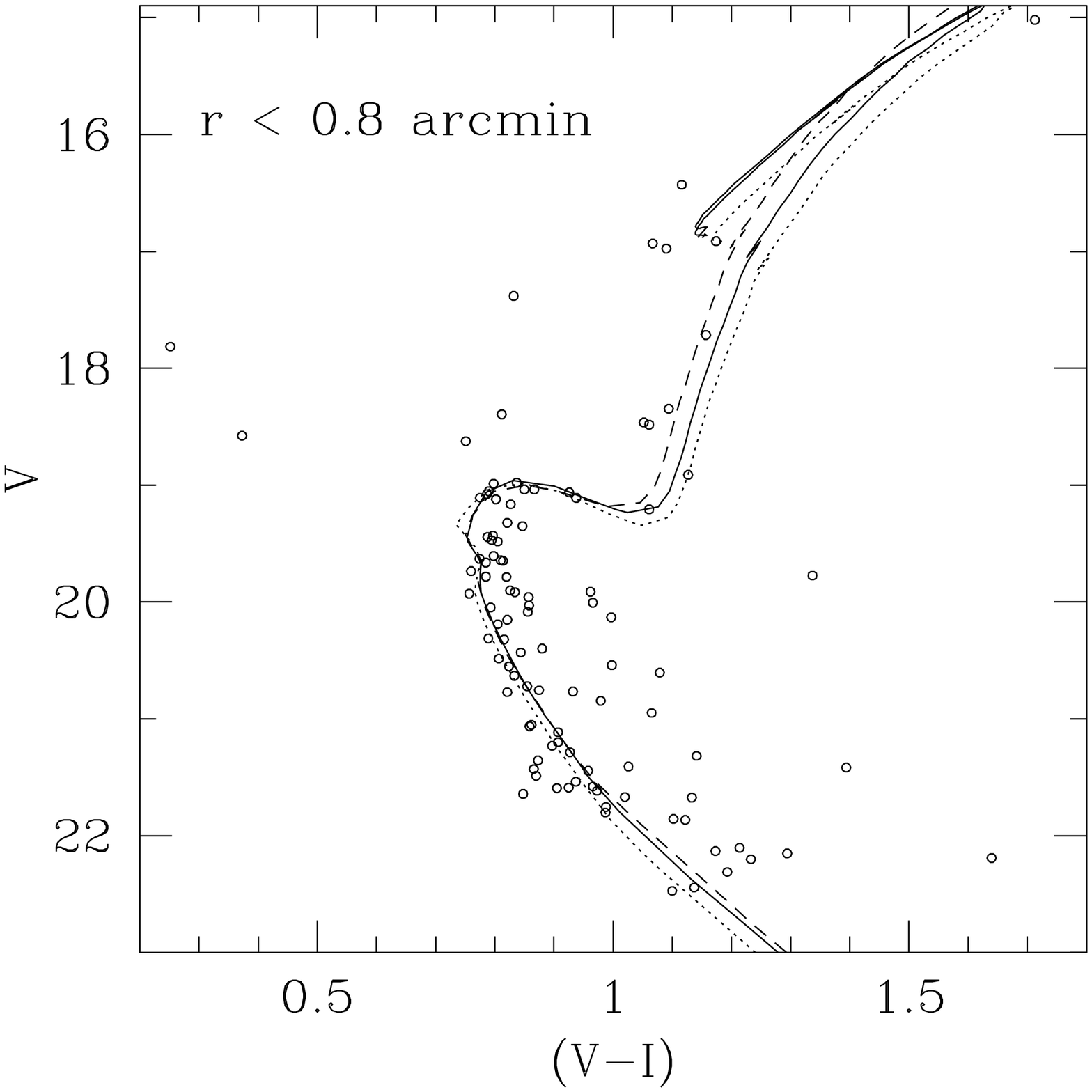,width=\columnwidth}} 
\caption{Age uncertainty for Saurer A
(only stars within 0.8  arcmin from the cluster center
are considered).
The {\bf solid line} is a 5 Gyr isochrone
for Z=0.008 metallicity, the {\bf dashed line} 
and {\bf dotted line} are
6.0 and 4.0 Gyr isochrones for the same metallicity,
respectively. See tex for more details}
\end{figure}

\subsection{The Colour-Magnitude Diagram} 
In Fig.~6 we present the CMDs of Saurer A (left panel) 
and the comparison field (right panel).
The two CMDs look like quite similar, and there is no
signature of a star cluster by considering the stars
in the two fields altogether. The only significant 
difference is the blue edge of the Main Sequence between
19 and 22 mag, which is much more populated in the
cluster than in the field, and which can be interpreted as
a cluster sequence.
To clarify this point, we consider only the stars in a 
circle 1.7 arcmin in radius (somewhat larger than the cluster radius)
wide centered in the cluster center,
and we  compare the derived CMD with its counterpart
in the field. These regions have been selected to match
the circles in Fig.~1 in FP02.\\

\noindent
The result is  shown in Fig.~7, from which
we see that actually a cluster exists, and the CMD
looks like that one from FP02, 
although we have more stars, and
the MS is 1 mag more extended. The exact shape of the CMD
is however difficult to understand, 
at the point we are not able to
distinguish neither a Red Giant Branch nor a clump.\\
In particular the detection of a clump at $V$=16.7,
$(V-I)$ = 1.15 by FP02 is suspicious, provided the almost
similar stars distribution of the cluster and the field
in the upper part of the CMD.
It is therefore clear that from this CMD it is not possible 
to derive reasonable estimates of the cluster parameters.\\

\noindent
To better deal with field stars contamination
we then consider only the stars confined within 0.8 arcmin
from the cluster center. (see Fig.~8)
Here the cluster appears very nicely and the contamination
of foreground stars is negligible (see also the field
CMD in the right panel).
The MS turn off (TO) is located at $V$=19.0, $(V-I)$=0.80,
as in FP02. The MS extends for 3.5 mag, and shows 
two probable gaps at $V$=19.25 and $V$=21.0
A sequence of binary is also visible red-ward the MS.
There are still a few interlopers, but the upper
part of the CMD appears now sufficiently clear,
although poorly populated. It is in fact possible
to see some RGB stars and a probable clump populated by 3 stars
at $V$=16.9, $(V-I)$=1.1 .

\begin{figure}  
\centerline{\psfig{file=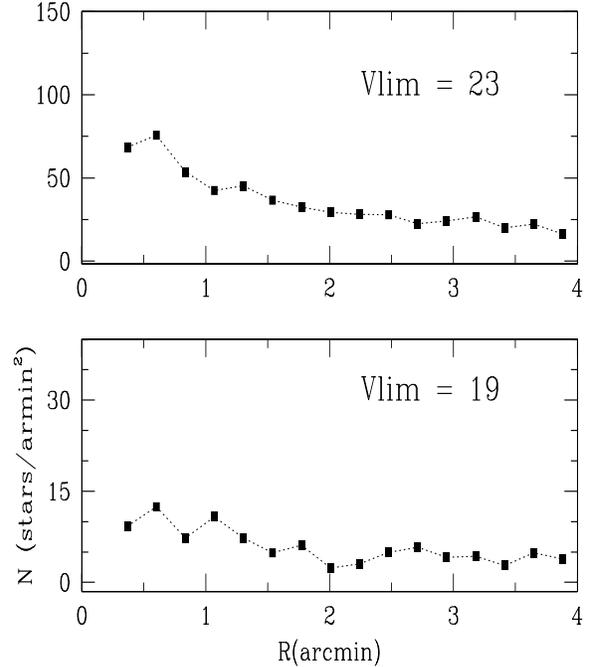,width=8cm,height=11cm}} 
\caption{Star counts in the region of Saurer B as a function
of the V magnitude.} 
\label{SaurerA} 
\end{figure}

\begin{figure*}  
\centerline{\psfig{file=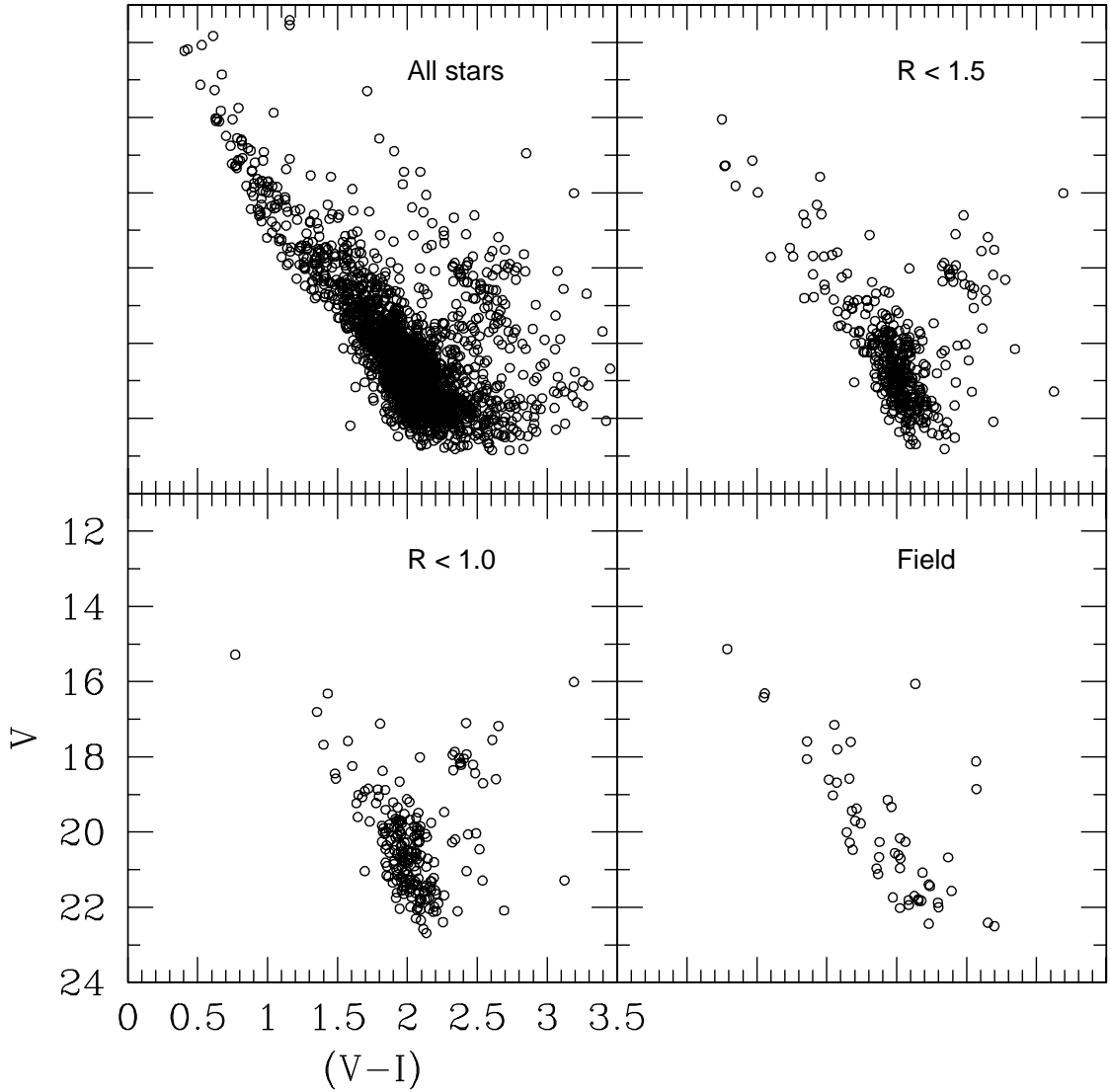,width=16cm,height=16cm}} 
\caption{{\bf Left upper panel:} CMD of Saurer B
for all the detected stars.
{\bf Right upper panel:} CMD of Saurer B for all the stars
within the cluster radius. {\bf Left lower panel:} CMD of Saurer B
for all the stars within 1.0 arcmin from the cluster center.
{\bf Right lower panel:} CMD of the stars located in a ring
with the same area as the cluster central part, and meant to
represent the field population.} 
\end{figure*} 
 
\subsection{Basic parameters}
We derive Saurer A fundamental properties by comparing
the CMD with theoretical isochrones from Girardi et al. (2000).
The choice of this set of models is motivated by the need
to keep the age, distance and reddening determinations
in the same scale of a series of previous papers on the same subject
(see for instance Carraro et al. 2002 and references therein).
This is quite a basic constraint, especially
when old open clusters are used as tracers of the Galactic
disk properties and evolution (see the discussion in Carraro
et al. 1998), where the homogeneity of the sample
is a fundamental need.\\

\noindent
In the following analysis
we are going to consider only the stars within 48 arcsecs.
The best fit isochrone solution is shown in Fig.~9,
where we over-imposed a 5 Gyr isochrone for 
Z = 0.008 metallicity. We have tried several combinations
of ages and metallicities, but none provides a reasonable
fit like this.\\ 
For the sake of illustration we over-plot in Fig.~9
also the solar metallicity isochrone which provides the most
reasonable fit (dashed line). This isochrone is for an age of 6.3 Gyrs,
and has been shifted by E$(B-V)$=0.09
and $(m-M_V)$=15.6. However, while the TO is nicely reproduced,
both the MS and the most evolved region of the CMD are poorly
accounted for.\\ 
To better bracket the metal content of the cluster we also
over-imposed the Z=0.004 isochrone which better fits
the data, although this metallity would be quite low
and unexpected for population I objects. The best fit
(dotted line in Fig.~9) is obtained for an age of 4.5
Gyrs, a reddening E$(B-V)$=0.20, and a distance modulus
$(m-M_V)$=16.2 . The fit is good also in this case, the MS
is actually a bit bluer, and in principle one could
increase a bit the age to lower the distance between
the TO and the RGB bottom, but in this way the clump would
become too bright. Notice (see Fig.~9) that the theoretical
clump is already brighter than the oserved one at 4.5 Gyrs.\\

\noindent
Therefore we opted for a half solar metal abundance.
The fit with the Z=0.008 isochrone
has been obtained by shifting the isochrone
by E$(V-I)$ = 0.18 and $(m-M_V)$ = 16.0 . These values are in nice agreement
with FB02. To get an estimate of the age uncertainty we have
over-imposed to the data  a younger and older isochrone 
for Z=0.008 metallicity. The result is shown in Fig.~10.
In this figure the solid line is the 5 Gyr best fit isochrone,
whereas the dotted line and the dashed line are a 4 and 6 Gyr 
isochrones, respectively. By keeping the isochrone close
to the cluster TO, the evolved region shows what one expects,
namely a bluer RGB at older ages, and a dimmer clump
at younger ages. 
From this figure we estimate an age uncertainty
less than 1 Gyr.\\

\noindent
FB02 uses the clump as distance indicator, by assuming
that the clump really exists and that its position does
not depend neither on age nor on metallicity.
The fact that the clump can be used as distance indicator
is quite well known. 
However we restrain to use it mainly because 
- as shown by Girardi \&Salaris (2001) - 
the clump position is a function of age
and metallicity. This latter parameter in particular
cannot be robustly constrained with the available data.

\noindent
We get an absoulte distance modulus $(m-M)_0$ =15.6.
As a consequence, 
Saurer A turns out to be 13.2 kpc far from the Sun,
and  by adopting $R_{\odot}$=8.5 kpc, its rectangular 
coordinates are: X = 19.3, Y= -7.4
and Z = 1.7 kpc, respectively. 
Therefore Saurer A is the most peripheral
open cluster to date, and lies very high on the Galactic plane
for an open cluster.\\

\noindent
In conclusion, from our photometry we better constrain Saurer A
basic parameters. In particular we refine the size and the age
of the cluster, we suggest that it is metal poor, and 
basically confirm the reddening and distance already 
found by FB02. As for the age, this is not an unexpected result.
It is very well known that MAI (Janes \& Phelps 1994)
tends to over-estimate the age of a cluster, and can be
used only as a qualitative indication of the relative
age between two or more clusters (see the discussion
in Carraro et al. 1999 and FB02).

\section{The open cluster Saurer B}

\subsection{Cluster radius}
Saurer B appears as a faint concentration of stars as 
Saurer A (see Fig.~2), although somewhat more extended and loose.
In order to infer a more robust estimate of the cluster radius, we performed
star counts by using our CCD data (2200 stars).
We derived the surface stellar density by performing star counts
in concentric rings  15 arcsecs  in size 
around the visual cluster center,
and then by dividing by their respective surfaces.
In this case we do not have a comparison field, so it is more
cumbersome to derive a firm estimate. However, by looking at Fig.~11,
one can conclude along the same vein of the discussion for Saurer A
that the cluster emerges over the field as a group of faint stars,
and that the cluster radius is $\approx$ 1.5 arcmin, a value which
confirms the visual inspection of Fig.~2.

\subsection{The Color-Magnitude Diagram}
In Fig.~12 we present various CMDs of Saurer B as a function
of the distance from the cluster center. In the upper left
panel the CMD for all the detected stars is shown. Here
the cluster is barely visible and the CMD is dominated by the
MS of the Galactic disk population. At odds with Saurer A,
this cluster is in fact located quite low in the Galactic disk.
Having estimated a radius of $\approx$ 1.5 arcmin, we present
also the CMD for the star in this region (upper right panel),
which basically shows the same features of the previous CMD,
although an important fact can be noticed:
while the blue part of the CMD in this case becomes narrower
and better defined, the red part does not change too much,
and the evidence appears of the possible presence
of a RGB clump.\\
To better probe the cluster population we show in the lower
left panel the stars enclosed within 1.0 arcmin from the cluster
center. Here we see a nice MS, although significantly wide,
and a well populated RGB clump. This is confirmed
also by the CMD in the lower right panel, which comprises
the same area as in the previous panel, and it is meant to
represent the field stars population. In this CMD there
is no clump at all.
The cluster TO is located at $V$ = 20.0, $(V-I)$ = 1.8,
while the clump is centered at $V$=18.2, $(V-I)$=2.4.
The width of the MS is much probably not due to photometric
errors (see Fig.~5, upper panel) which at $V \approx$ 22 amounts at
less than 0.1 mag in color. Therefore we suggest the MS is that wide
due to other two possible reasons, which however
we are not possible to quantify
with the present data: a binary population and some differential reddening
across the custer.\\

\noindent
In conclusion Saurer B exhibits all the features of an intermediate
age open cluster (Carraro et al. 1999), and resembles
very much the CMD of clusters like NGC~2158 (Carraro et al. 2002)
and NGC~7789 (Girardi et al. 2000).

\begin{figure} 
\centerline{\psfig{file=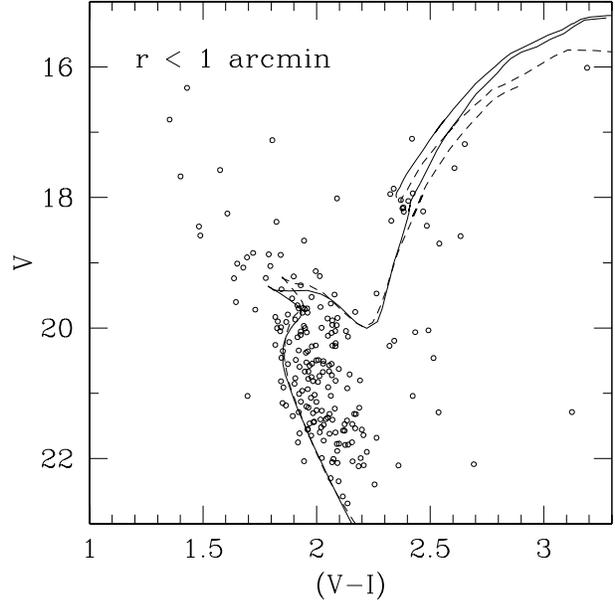,width=\columnwidth}} 
\caption{Saurer B stars within 1 arcmin from the cluster center
 in the $V$ vs.\ $V-I$
diagram, as compared to Girardi et al. \ (2000) isochrone of age 
$1.8\times10^9$ yr 
(solid line), for the metallicity $Z=0.008$. A distance 
modulus of $(m-M)_0=14.10$ mag, and a colour excess of E$(V-I)=1.38$ mag, 
have been derived. The dashed line is on the other hand a 2.2
Gyrs solar metallicity
isochrone shifted by E$(V-I)=1.30$  and $(m-M)_0=14.08$} 
\end{figure} 
 
\begin{figure}  
\centerline{\psfig{file=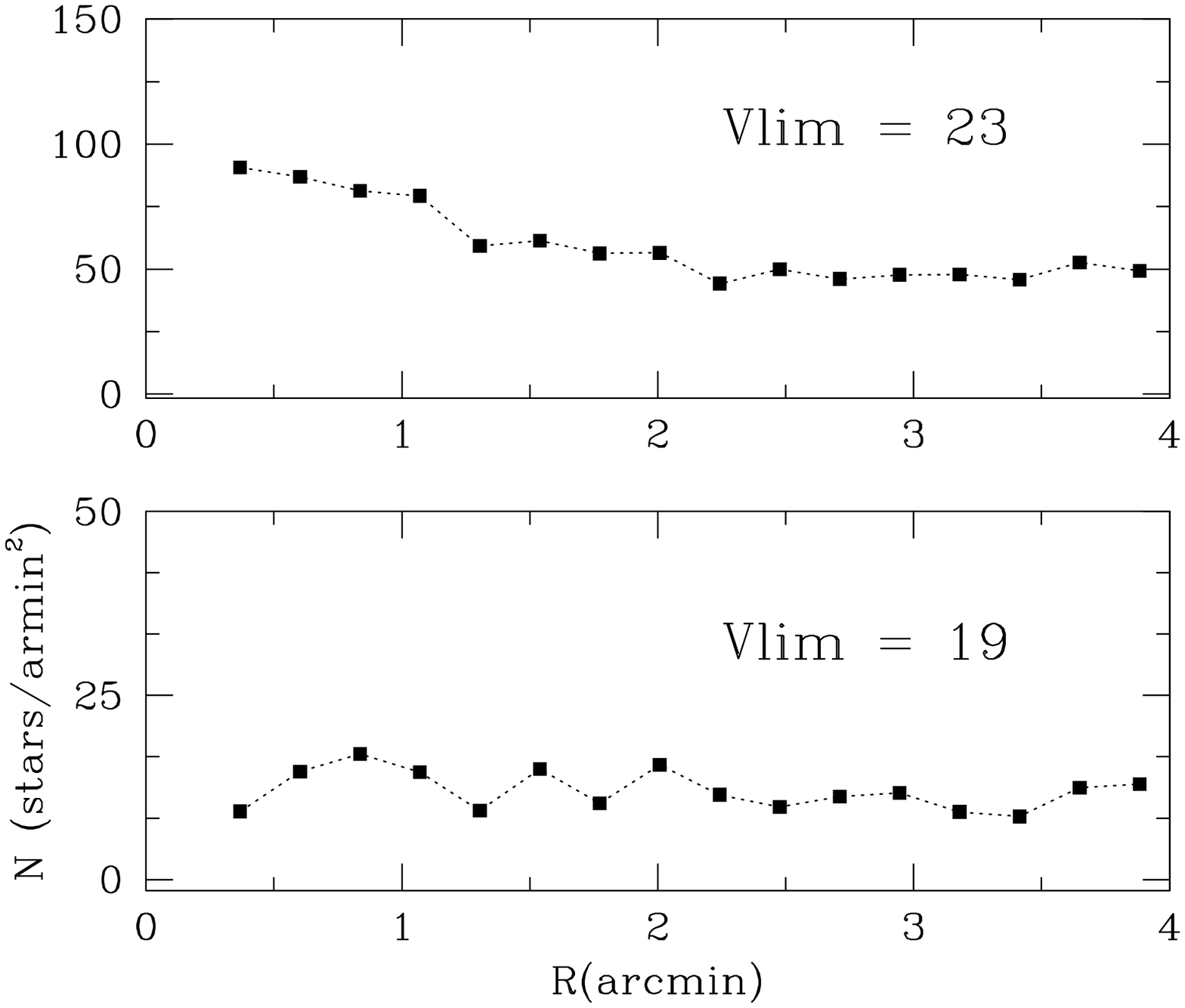,width=8cm,height=11cm}} 
\caption{Star counts in the region of Saurer C as a function
of the V magnitude.} 
\end{figure} 
 
\subsection{Basic parameters}
We derive Saurer B fundamental parameters in the same way as for Saurer A.
In Fig.~13 we plot all the stars lying within 1 arcmin from the adopted 
cluster center, and we show the best fit isochrone solution.
Again we use a Z=0.008 isochrone (solid line), 
but for the age of 1.8 Gyrs,
which nicely fits both the TO region and the RGB clump.
The fit has been obtained by shifting the isochrone by
E$(V-I)$=1.38 and $(m-M_V)$=17.4, and the corrected distance modulus turns
out to be  $(m-M_V)_0$=14.1.\\
In order to derive an estimate of the cluster metal content,
also in this case we tried a fit with a solar
metallicity isochrone (dashed line), and find an age of 2.2 Gyrs,
a reddening E$(V-I)$=1.30, an apparent distance modulus $(m-M_V)$=17.2,
and corrected distance modulus turns
out to be  $(m-M_V)_0$=14.08.\\
The quality of the CMD - in particular the region
of the TO is probably still affected by some contamination -
does not allow us to firmly establish the
metallicity of the clusters and, indeed, the derived parameters
are pretty similar.

\noindent
In this case, however, the cluster clump is clearly visible,
and  placed  at V=18.2. Therefore
we derive (Girardi \& Salaris 2001) that for an age of 1.8 Gyrs
and a metallicity of Z=0.008, the absolute clump magnitude
is $V$ = 0.542, and therefore $(m-M_V)$ = 17.7.
In the case of solar abundance, the clump magnitude at 2.2 Gyrs
is $V$ = 0.578, hence $(m-M_V)$ = 17.6.
Also with this method we basically obtain the same distance
modulus, stressing again our inability to discriminate between
different metallicity models.

\noindent
As a consequence of these outcomes, Saurer B is placed  6.6 kpc far from the Sun,
and its rectangular coordinates are: X = 9.8, Y= -6.5
and Z = -0.1 kpc, respectively. \\

\noindent
To summarize,  with respect to the study of FB02,
we obtain a much larger distance from the Sun, and a 
significantly younger age, a result which confirms
the trend that the  MAI predicts systematically
older ages.

\begin{figure*}  
\centerline{\psfig{file=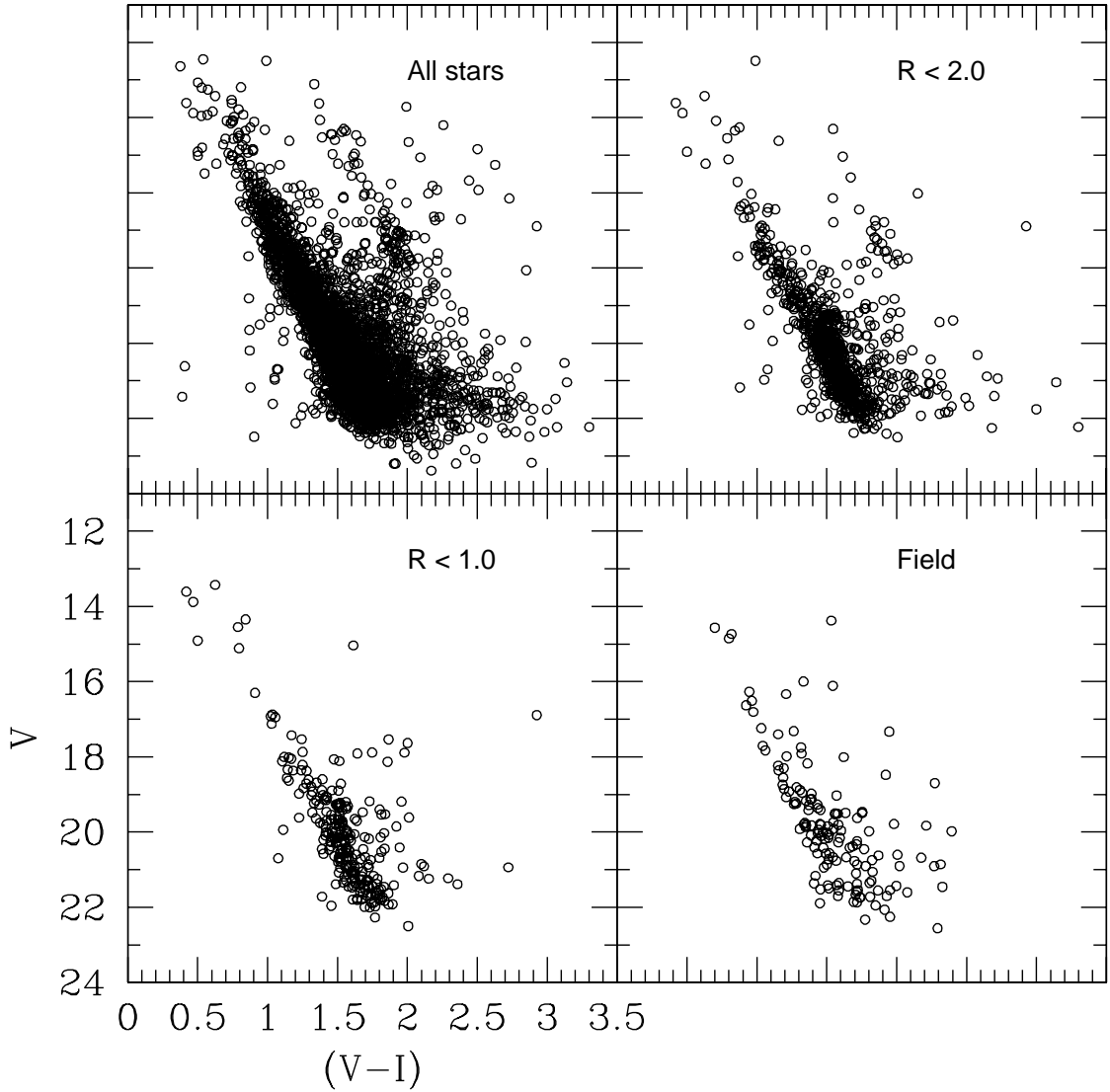,width=16cm,height=16cm}} 
\caption{{\bf Left upper panel:} CMD of Saurer C
for all the detected stars.
{\bf Right upper panel:} CMD of Saurer C for all the stars
within the cluster radius. {\bf Left lower panel:} CMD of Saurer C
for all the stars within 1.0 arcmin from the cluster center.
{\bf Right lower panel:} CMD of the stars outside 
the cluster radius, in an area equal to that one in the previous panel.} 
\end{figure*} 

\section{The open cluster Saurer C}

\subsection{Cluster radius}
Saurer C appears as a faint concentration of stars 
in a very rich Galactic field (see Fig.~3).
In order to achieve an   estimate of the cluster radius, we performed
star counts by using our CCD data (4500 stars).
We derived the surface stellar density by performing star counts
in concentric rings  half an arcmin in size 
around the visual cluster center,
and then by dividing by their respective surfaces.
The results are shown in Fig.~14.
In this case  we find an over-density of faint stars up to a radius
of $\approx$ 2 arcmin, while the bright stars profile
keeps flat, showing that the cluster does not contain
a significant amount of bright stars.
By combining together the shape of the density profile
and the appearance of the cluster in Fig.~3, we suggest that the
cluster radius is around 2 arcmin.

\begin{table*}
\caption{{}Basic data of the observed objects.}
\fontsize{8} {10pt}\selectfont
\begin{tabular}{ccccccccc}
\hline
\multicolumn{1}{c} {$Name$} &
\multicolumn{1}{c} {$Radius(arcmin)$} &
\multicolumn{1}{c} {$E(V-I)$}  &
\multicolumn{1}{c} {$E(B-V)$}  &
\multicolumn{1}{c} {$(m-M)_0$} &
\multicolumn{1}{c} {$X(kpc)$} &
\multicolumn{1}{c} {$Y(kpc)$} &
\multicolumn{1}{c} {$Z(kpc)$} &
\multicolumn{1}{c} {$Age(Gyr)$} \\
\hline
Saurer A & 1.3 & 0.18 & 0.14 & 15.6 & 19.3 & -7.4 &  1.6 & 5.0$\pm$1.0\\
Saurer B & 1.5 & 1.30-1.38 & 1.05-1.10 & 14.1 &  9.8 & -6.6 & -0.1 & 1.8-2.2\\     
Saurer C & 2.0 & 0.98 & 0.76 & 14.9 & 11.0 & -9.3 &  0.5 & $\approx$2.0\\     
\hline
\end{tabular}
\end{table*}

\subsection{The Color-Magnitude Diagram}
In Fig.~15 we present several CMDs of Saurer C as a function
of the distance from the cluster center. In the upper left
panel the CMD for all the detected stars is shown. Here
there is no cluster appearance  and the CMD is dominated by the
Galactic disk field stars population.
If we consider only the stars located inside the cluster radius
(upper right panel), the situation does not change too much, the only
improvement being that the red part of the CMD is better defined,
and a RGB clump seems to be present.
However, if we consider the more central part of the cluster
(lower left panel), the clump becomes poorly
populated, rendering very difficult
the interpretation of the CMD. Fortunately, when considering
a field area (lower right panel)
of the same size of the cluster central part,
we find that there is no hint for a clump, thus making us
more confident with the interpretation of the CMD in the
lower left panel.

\subsection{Basic parameters}
Although the field stars contamination is very severe in the field
of Saurer C, 
we still tried to find an isochrone solution, which of course has to
be considered preliminary. The result is shown in Fig.~16,
where we plot all the stars within 1 arcmin from the cluster radius.
The fit has been obtained by shifting the isochrone by
E$(V-I)$=0.98 and $(m-M_V)$=17.2, and the corrected distance modulus turns
out to be  $(m-M_V)_0$=14.9.
Since the clump is placed  at $V$=18.0,
 we derive (Girardi \& Salaris 2001) that for an age of 1.8 Gyrs
and a metallicity of Z=0.008, the absolute clump magnitude
is $V$ = 0.540, and therefore $(m-M_V)$ = 17.46, in fine agreement
with that derived from the isochrone fitting\\
\noindent
As a consequence, Saurer C is placed  9.6  kpc far from the Sun,
and its rectangular coordinates are: X = 11.0, Y= -9.3
and Z = 0.5 kpc, respectively. \\

\begin{figure} 
\centerline{\psfig{file=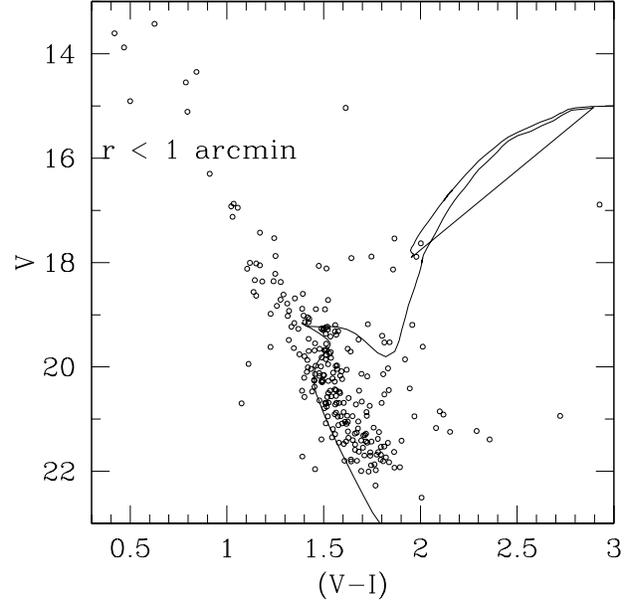,width=\columnwidth}} 
\caption{Saurer C star within 1 arcmin from the cluster center
in the $V$ vs.\ $V-I$
diagram, as compared to Girardi et al. \ (2000) isochrone of age 
$2.0\times10^9$ yr 
(solid line), for the metallicity $Z=0.008$. A distance 
modulus of $(m-M)_0=14.90$ mag, and a colour excess of E$(V-I)=0.98$ mag, 
have been derived.} 
\end{figure} 
 
\section{Conclusions}
We have presented deep CCD $VI$ photometry study of the 
old open clusters Saurer A, B and C. The CMDs we derive allow us to 
constrain quite well the cluster basic parameters, which are listed in 
Table~3.
In summary, we find that:
 
\begin{description} 
\item $\bullet$ Saurer A is an M~67 like old open cluster and it is the most 
distant open cluster to date; it would be of extreme 
interest to have a spectroscopic confirmation
of its metal abundance. RGB stars at $16 \leq V \leq18$ are indeed 
easily affordable with present day 8m class telescopes; 
\item $\bullet$ Saurer B turns out to be a very reddened NGC~2158 like,  
intermediate-age open cluster;
\item $\bullet$ Saurer C is as well an intermediate-age open cluster,
but it remains a very difficult object due to the
heavy field stars contamination toward its direction.
\end{description}

\section*{Acknowledgements} 
We are pleased to thank John Willis for his kind introduction
at ESO NTT, and for useful suggestions during the reduction
of the data presented in this paper.
This study has been financed by the Italian Ministry of 
University, Scientific Research and Technology (MURST) and the Italian 
Space Agency (ASI), and made use of Simbad and WEBDA databases.

\end{document}